# Total ionizing dose test with DEPFET sensors for Athena's WFI

Valentin Emberger*[a], Michael Bonholzer[a], Johannes Müller-Seidlitz[a] and Robert Andritschke[a]
[a]Max Planck Institute for Extraterrestrial Physics (MPE), 85748 Garching, Germany

## ABSTRACT

The focal plane of Athena's WFI consists of spectroscopic single photon X-ray detectors that contain arrays of DEPFETs (DEpleted P-channel Field-Effect Transistor) as well as ASICs that are used for steering, readout and analog signal shaping. These components have to be examined regarding the effect of ionizing radiation. A Total Ionizing Dose (TID) test was done with prototype detector modules with 64×64 DEPFETs and one SWITCHER and VERITAS ASIC each. The current design of the WFI detector head features a proton shield equivalent to 4 cm of aluminum in order to prevent a strong increase of leakage current in the fully depleted 450 µm thick bulk of the sensor. This keeps the expected doses and dose rates during the nominal mission relatively low (~5 Gy). It is nevertheless important to study the current system in a dedicated TID test in order to exclude unforeseen effects and to study any radiation related changes that can have an effect on the very sensitive readout chain and the detector performance. The combination of low doses, low dose rates, low operating temperature (<-60°C) but high sensitivity on small changes of the threshold voltages represent somehow unusual boundary conditions in comparison to TID tests for standard radiation hard electronic components. Under these circumstances it was found beneficial to do the test in our own laboratory with an X-ray source in order to realize irradiation during nominal operation conditions. Furthermore, it facilitated to take annealing effects into account. Reasonably accurate dosimetry is achieved by measuring the X-ray spectrum and intensity with the device under test. After irradiation to a total dose of 14 Gy and subsequent annealing the threshold voltage of the DEPFETs were shifted by a mean value of 80 mV, the performance remained unchanged apart from a slight increase in readout noise by 10%.

**Keywords:** Athena WFI, DEPFET, X-ray detector, TID, total ionizing dose, threshold voltage shift, interface states

## 1. INTRODUCTION

Athena's Wide Field Imager (WFI) will perform imaging spectroscopy with near fano-limited energy resolution and high time resolution. It consists of back illuminated, fully depleted DEPFET active pixel sensors. The large detector array with more than 1 Mpixel in total consists of four large detectors with 512×512 pixels each. The size of the pixels is 130×130 (µm)$^2$ which results in a sensitive area of 177 cm$^2$. That enables a wide field of view of 40'×40' [1-3].

The fast detector (64×64 pixels) has a timing resolution of 80 µs. Both detectors are read out in rolling shutter mode, being steered by SWITCHER ASICs [2,4] that sequentially switch on the sensor rows. They also control the voltages necessary for clearing the charges stored in the pixel center (the so-called internal gate). Each channel of the sensor is connected to one input of a VERITAS readout ASIC [5] that is responsible for reading, shaping and multiplexing the analoge signals. The readout is done in the drain current readout configuration, where the source of the DEPFET is kept at a constant voltage and the VERITAS, that is connected to the drain of the DEPFET, senses the change of drain current due to the amount of signal electrons in the internal gate of the pixel.

An orbit around L1 is foreseen for Athena, thus the WFI instrument will be subject to ionizing radiation in the form of solar particles and galactic cosmic rays. In order to ensure the continued operation with very low noise readout (~3 e$^-$ ENC) and good energy resolution (<80 eV @ 1 keV) the focal plane has to be shielded by a 4 cm thick aluminum proton shield in all directions other than towards the mirror. Otherwise the 450µm thick bulk of the sensor would accumulate enough displacement damage to result in a prohibitively large increase of dark current. Due to this strong shielding the sensor is expected to receive only a moderate total ionizing dose (~5 Gy in the first 10 years of operation according to preliminary simulations with FASTRAD [6]).

A prototype detector module (see section 2) with a sensor from Athena's pre-flight production was used to study the effect of TID. In section 3 we describe the used radiation source and measurement setup. The definition of the required dose and dose rate, the determination of the corresponding flux and the dosimetry are detailed in section 4. In section 5 the effects caused by the irradiation are shown and all measurements for characterizing the detector before and after irradiation are described. The possible implications of radiation effects on the drain readout concept are discussed and an outlook of further necessary radiation tests is given.

---
*send correspondence to emberger@mpe.mpg.de

## 2. TEST SPECIMEN

The TID test was done with a prototype detector module with a DEPFET sensor from the Athena pre-flight production. In this production almost all devices on a wafer have the same pixel layout, including one large device (512×512 pixels) and several smaller devices with 64×64 pixels. The detector module is shown in Figure 1, the sensor contains 4096 independent DEPFET pixels whose irradiation response is studied individually.

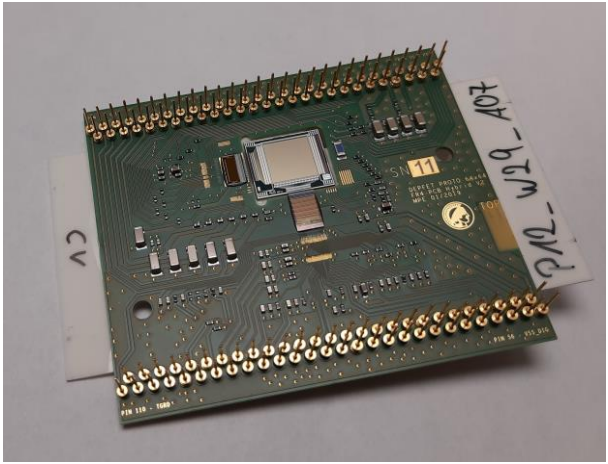
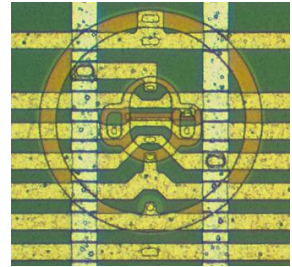

**Figure 1:** Left: detector module, the sensor (64×64 pixels) is glued to a ceramic carrier that is attached to a PCB with front end electronics. SWITCHER ASIC is left of the sensor, VERITAS below. Top: Pixel layout of the sensor, metal layers appear lighter (white/yellow) than polysilicon structures (orange).

## 3. RADIATION SOURCE AND TEST SETUP

As source of radiation a standard analytical X-ray tube in a Seifert SH 37/80 RÖ housing was used. It is supplied by a GE/Seifert ISO-DEBYEFLEX 3003. The tube is attached to a vacuum chamber in which the detector module is installed and cooled to the operating temperature of -60°C. The tube is equipped with a molybdenum target. The Mo-target was preferred over a Fe- or Cu-target in order to obtain a higher energy of the characteristic lines. This leads to a more uniform depth profile of the absorbed radiation in the device under test. The tube is operated with a high voltage of 30 kV, the intensity is adjusted by setting the anode current between 2 and 50 mA. A filter wheel is in-between the tube and the detector where different filters can be inserted. Relatively thick filters (~100µm) of zirconium and molybdenum are used which lead to a spectrum that is strongly dominated by the Mo-$K_\alpha$ line. This makes the calculation of the dose less susceptible to systematic errors (see section 4.3).

Even though the use of X-rays is relatively uncommon for the testing of radiation hardness testing, it was found beneficial in this case for several reasons. A big advantage is the availability in our own laboratory, which facilitated to maintain all operation conditions as they are specified for the flight model. The detector was in vacuum, cooled to -60°C, fully biased and operated during all irradiations. Performance measurements as well as electrical characterization could be carried out in-situ in-between and after the irradiations. The behavior during annealing could be monitored for several weeks without even shortly interrupting the biasing of the detector and maintaining full control of the detector temperature. A second advantage stems from the fact that the device under test is a spectroscopic X-ray detector. Using an X-ray source allows to use the specimen for the dosimetry. Thus the spectrum, intensity and spatial homogeneity of the radiation could be monitored at all times exactly at the location of interest. Because the sensitive components (ASICS and sensor) are completely unpackaged the issue of low and strongly energy dependent penetration of X-rays through the packaging is irrelevant here.

The detector module is integrated in the test chamber with the front side of the sensor facing towards the X-ray source (the front side is the structured side of the sensor). Only this orientation allows the simultaneous irradiation of sensor and ASICs as the latter are glued on a relatively thick PCB. The distance between filters and sensor is greater than 1 m and the optical path is free of obstacles which makes the irradiation completely homogeneous. A movable $^{55}$Fe-source for performance measurements is on the opposite side of the sensor in order to illuminate the back side as in nominal operation. The test chamber is evacuated to a pressure below $10^{-6}$ mbar and the specimen can be cooled with a Ricor K535 stirling cooler to the operational temperature of -60°C.

## 4. DOSIMETRY

### 4.1. Expected Dose

The interaction of the WFI instrument with the expected radiation environment at L2 was studied in a number of dedicated simulations [7-9]. The aim of these studies was to improve the instrument design with respect to the non-X-ray background. With the mass model and software setup that were developed for these simulations also the expected dose could be calculated. For a ten-year period starting in 2028 the TID at a confidence level of 95% was found to be $4.97 \pm 0.1$ Gy(Si). The planned orbit for Athena has meanwhile changed from L2 to L1, but for the sake of this test the radiation environment at L2 and L1 can be considered equal. The simulations show that the main dose will be caused by protons in the 10 to 200 MeV range with a relevant contribution of secondary electrons. This information is useful to determine the conversion factor that accounts for the different efficiency in terms of radiation damage by different radiation sources (section 4.2). To account for the uncertainties associated to the conversion factor, to the preliminary mass model used in the simulation and the launch date etc. a radiation design margin of 200% was chosen. Thus it was specified to irradiate the specimen to 15 Gy in three steps.

### 4.2. Conversion factor for X-ray source

In this test X-rays from a tube with Mo-target are used. The standards for radiation testing are, however, most often referring to the use of $^{60}$Co γ-radiation. On the other hand, the simulations show that the dominant TID contribution is likely to originate from protons in the 100 MeV range. Thus, in the following it will be shown how the radiation damage that results from the used X-rays is related to that from $^{60}$Co γ-radiation and 100 MeV protons.

Several publications suggest the use of low energy X-rays for radiation hardness testing, especially if protons are expected to be the main source of radiation damage [10-12]. The reason for this is that the recombination probability (and thus the initial charge yield) for protons in this energy range is closer to that of low energy X-rays than to that of $^{60}$Co γ-radiation. According to [12] the radiation damage from high energy protons for a given dose is only 65% -85% that of $^{60}$Co or ~1 MeV electrons at the same dose. But it has to be taken into account that in these papers 100 MeV protons are compared to X-rays from a Bremsstrahlung continuum with a maximum at 10 keV. The secondary electrons from Mo-K$_\alpha$ photons (17.4 keV) have a lower LET (Linear Energy Transfer) than those from 10 keV X-rays, resulting in a higher charge yield. In [13] charge yields for different photon energies are given, but these depend also on the field strength. According to device simulations the field strength in the gate oxide of a linear gate DEPFET is on the order of $3 \cdot 10^5$ V/cm (Figure 2). For $3 \cdot 10^5$ V/cm the initial charge yield is shown to be approximately 30% higher at 17.4 keV (Mo-K) than at 8 keV (Cu-K).

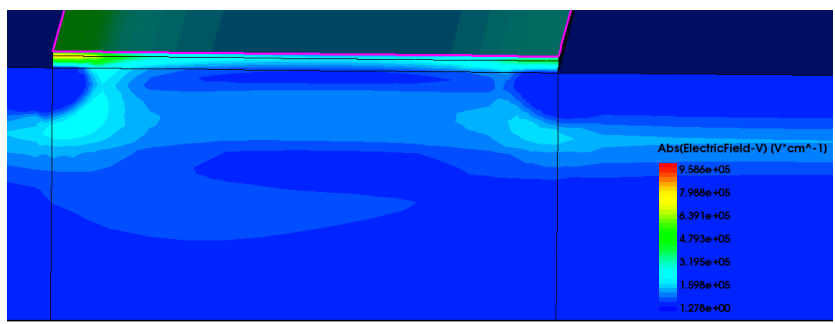

**Figure 2:** Simulation result for the electric field in the DEPFET sensor for a gate voltage that switches the DEPFET current off. The off state is most relevant as the device is only 1/64$^{th}$ of the time in the on state. The two uppermost layers are silicon nitride and silicon oxide

It can be concluded that the initial charge yield of the used X-rays has to be between that of $^{60}$Co γ-radiation and 100 MeV protons. As a precise determination of the initial charge yield is difficult and depends on many factors it was decided that no conversion factor will be applied. The uncertainties connected to the initial charge yield are reflected in a large radiation design margin.

### 4.3. Signal to dose ratio

In this TID test the dosimetry is done with the device under test. For this it is necessary to know the relation of the signal that is recorded by the DEPFET detector module to the applied dose. This relation depends on the spectral shape of the incoming X-rays (which in turn depends on the high voltage setting and the used filters) but it does not depend on the intensity that can be set by adjusting the anode current. In order to model the spectral shape of the incoming X-rays a spectrum with low intensity is recorded in order to reduce the energy pile-up which distorts the spectral shape and cannot easily be included in the model.

The following settings were applied in order to record an almost pile-up free spectrum:
- Frame time of 20 μs (such a short frame time is realized by readout time for one line of 2.5 μs 4444and a window mode where only 8 of the 64 rows are read out and the rest is skipped)
- X-ray source: HV= 30 keV, I = 2mA (lowest possible current)
- Thick filters: 86 μm Zr + 95 μm Mo

One million frames were recorded; the resulting spectrum is distributed into bins of 20 eV width (shown in Figure 3). It is calibrated with the standard "Offline-Analysis-Tool" [14] with the use of the Mo-$K_\alpha$ line at 17445.6 eV. The spectrum is modelled with a function that is comprised by the following components:
- Two Gaussians for the $K_\alpha$ and $K_\beta$ lines
- Two functions that represent the so-called flat shelf that is mainly caused by energy misfits and energy losses. Energy misfits are the result of photons hitting a pixel during certain periods of the readout process. Their energy is only partially registered. These two functions are based on the cumulative distribution functions of the Gaussians of $K_\alpha$ and $K_\beta$ line.
- A model for the Bremsstrahlung. It is based on the Kramer formula multiplied with the transmission of the used filters. To account for the smearing of the absorption edges by the detector response it is convoluted with a Gaussian.

The model does not include pile-up. As the shape of the Bremsstrahlung spectrum depends critically on the transmission of the filters, the thicknesses of the filters are also free parameters in the fitting routine.

With the results from the fit the spectral shape of the input spectrum can be reconstructed (Figure 4). The reconstructed input spectrum consists of four monochromatic lines ($K_{\alpha1/2}$ and $K_{\beta1/2}$) and Bremsstrahlung. The α/β intensity ratio is a result from the fit, the intensity ratios of the subshell lines within $K_\alpha$ and $K_\beta$ are tabulated values. This spectrum is calculated by considering the spectral redistribution by the detector response, like the broadening of the lines due to Fano noise, etc. But it does not yet consider the limited Quantum Efficiency (QE) due to the finite thickness of the detector which is done in the next step.

As a model for the QE the absorption in 450 μm silicon is used (Equation 2). Because the position where the applied dose is specified is just at the Si/SiO$_2$ interface the SiO$_2$-, polysilicon-, Al- and Benzocyclobuten(BCB)-layers on the front side must not be considered here because they are present both during the measurement and the irradiation.

For each bin in the reconstructed input spectrum with height H(E), the dose d(E) in a 1 μm thick layer of silicon at the surface of the sensor is then given by

$$d(E) = H(E) \cdot \frac{1}{QE} \cdot \left(1 - e^{-t/\mu(E)}\right) \cdot E \cdot \frac{1.694 \cdot 10^{-19} \frac{J}{eV}}{\rho \cdot A_{sen} \cdot t} \qquad (1)$$

where

$$QE = \left(1 - e^{-450/\mu(E)}\right) \qquad (2)$$

and
E: Energy in eV  
H: Number of counts at energy E  
μ: Attenuation length in silicon  
ρ: Density of silicon  
$A_{sen}$: Sensitive sensor area  
t: Thickness = 1μm.

The thickness is arbitrarily chosen to be 1 μm, but as the depth profile of the absorption is very uniform this does not introduce a significant error. This is true for any chosen thickness that is much smaller than the attenuation length in the regarded energy range (the attenuation length in Si is 230μm at 12 keV and 1 mm at 20 keV). The sum of all doses from all bins in the spectrum yields a total dose of 1.97 mGy(Si) (~0.2 rad). This dose can be set in relation to the signal that is registered by the detector. This is done by integration of the unmodified spectrum. The total signal during the measurement was 148.9 MeV per pixel. Thus, the relation of signal to dose is calculated to be 13.22 mGy/(GeV/pix).

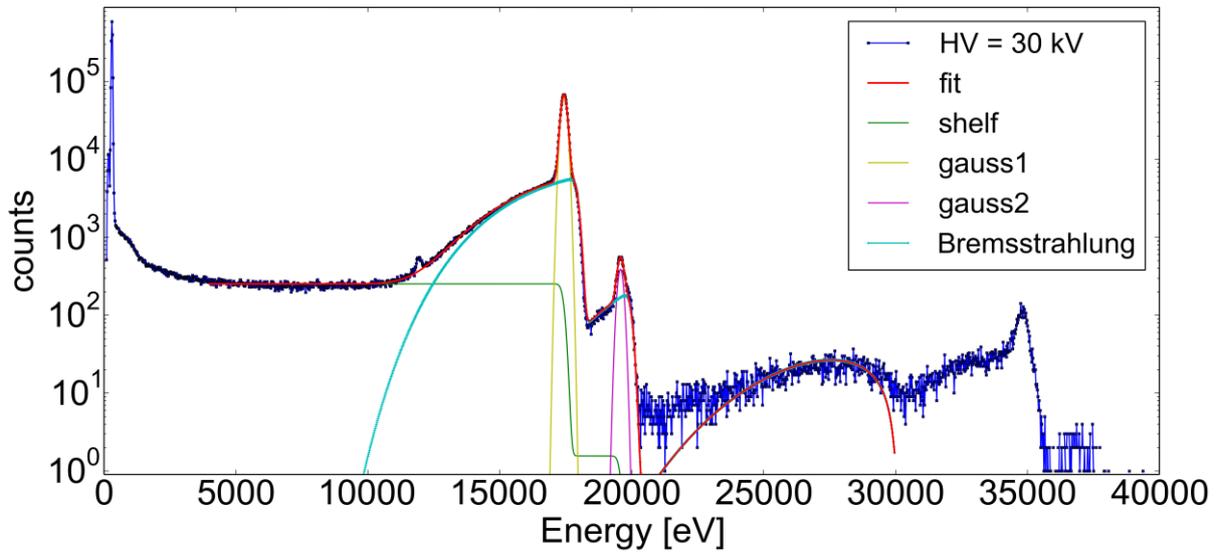

**Figure 3:** Measured spectrum of the X-ray source at a high voltage of 30 kV (blue). Also shown is the fit (red) and the different components of the fit function.

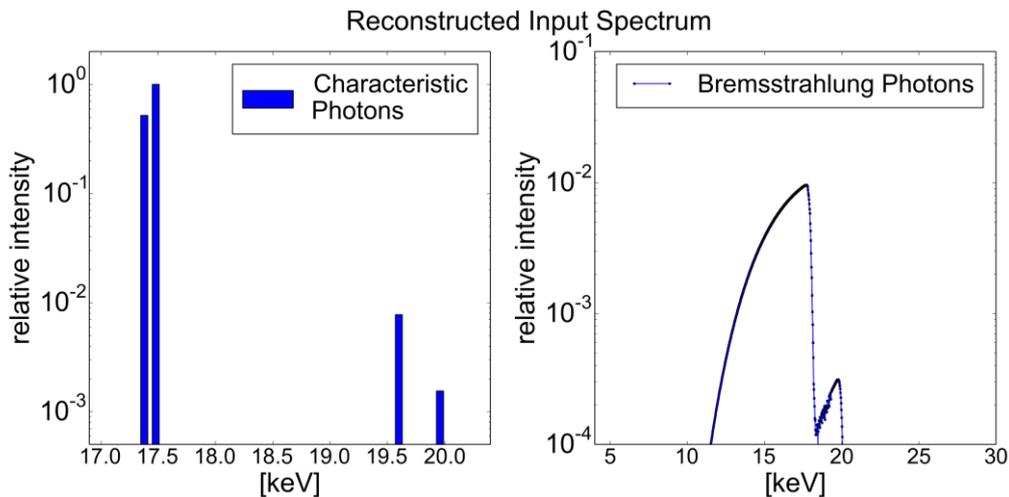

**Figure 4:** Spectrum of the X-ray source, reconstructed from the measurement in Fig. 3.

### 4.4. Flux (dose rate) measurements

The signal to dose relation was used to determine the dose rate during the irradiation. Three thousand frames were recorded in irregular intervals. The signal in each pixel is summed up and is converted in a dose rate with the use of the above given conversion factor and the measurement time. Figure 5 shows the dose rate during the three irradiation rounds. During the first irradiation the dose rate was considerably below the targeted 1 Gy/h; the anode current was adjusted for the following two rounds. In general, the rates were relatively stable. The mean dose rates are 786 mGy/h, 1008 mGy/h and 1010 mGy/h.

The irradiation time was 5 h for all three rounds. Consequently, the applied doses are: 1st Irradiation: 3.93 Gy, 2nd Irradiation: 5.04 Gy, 3rd Irradiation: 5.05 Gy. It is estimated that these doses are accurate within ± 10%.

## 5. RESULTS

The detector module was irradiated in three steps to a total dose of 14 Gy. Subsequently the device was annealed for 2 weeks at the operating temperature of -60 °C, and another 4 weeks at 20 °C. Bias was applied uninterrupted during all irradiations and during annealing without interruption. A number of characterization measurements were carried out before, during and after the irradiations. The energy resolution at 5.9 keV was determined from a Mn-$K_\alpha$ spectrum and the dark frame noise was measured.

Table 1 summarizes the measurement results. During the irradiations and the cold annealing the energy resolution and noise remained unchanged. The change in gain reflects the shift of the threshold voltages, as these lead to a lower drain current and thus to a lower gain. It can be seen that the threshold voltages are shifted after every irradiation and then further shifted during the annealing.

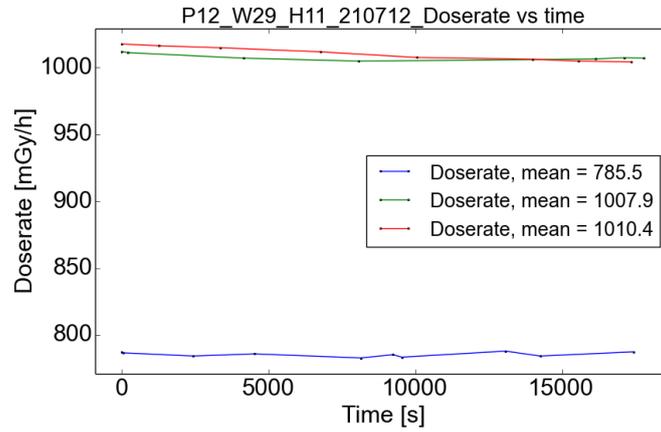

**Figure 5:** Measurements of the dose rate during the three rounds of irradiation

This indicates the build-up of positively charged interface traps in the transistor channel. These further push the DEPFET's characteristic - as they are p-channel devices - in the direction of the off state. They act in addition to the trapped holes in the $SiO_2$ on the other side of the interface. After the annealing at room temperature, the gate voltage of the DEPFETs was adjusted by 90 mV in order to get the drain current to the initial value. With the adjusted voltage the gain was measured to be very close to the initial value, as it is shown in the last line of Table 1.

The readout noise remained constant during irradiations and cold annealing but was increased by 10% after the annealing at room temperature. This noise increase could not be reversed by adjustment of the voltages; it might be caused by the formation of interface states.

In addition, the threshold voltages were measured pixel wise, a detailed description of the measurement method is given in [15]. The results are shown in Figure 6. The mean shift of the threshold voltage is 80 mV, which closely matches the value that was found by adjusting the gate volte. The lateral distribution of the shift is relatively uniform.

**Table 1:** Summary of the results of the performance measurements

|  | ID | Noise [adu] | Noise [e- enc] | FWHM [eV] | Gain [adu/eV] |
|---|---|---|---|---|---|
| Pre irradiation | 210705_02 | 9.70 | 2.5 | 129.7 | 1.016 |
| Post 1st Irradiation | 210705_23 | 9.69 | 2.6 | 129.3 | 1.007 |
| Post 2nd Irradiation | 210708_11 | 9.70 | 2.6 | 129.3 | 1.004 |
| 3rd Irradiation+ 1 week | 210719_01 | 9.69 | 2.6 | 129.4 | 1.005 |
| 3rd Irradiation+ 2 weeks | 210726_02 | 9.71 | 2.6 | 129.6 | 1.002 |
| Post RT annealing | 210823_02 | 10.78 | 2.9 | 131.0 | 0.994 |
| Post RT annealing $V_{gate}$ adjusted | 210901_02 | 10.71 | 2.8 | 131.0 | 1.013 |

## 6. CONCLUSION AND OUTLOOK

The WFI prototype detector has shown to be robust to withstand total ionizing doses of 14 Gy without significant changes in the performance. The mean observed threshold voltage shift is 80 mV. During biased annealing the readout noise and the threshold voltage shift increased which is attributed to the formation of interface traps.

The width of the initial distribution of the threshold voltages is ~300 mV. The lateral distribution of the threshold voltage shift does not show any obvious correlation to the map of the initial threshold voltage map. This indicates that the differences in threshold voltage between pixels are not amplified due to the irradiation. Hence, in the case of homogeneous irradiation the distribution has a similar width. This is important as it has a consequence on the spread of the individual pixel currents that must always be below a certain value so that all pixels stay within the input range of the first stage in the VERITAS readout ASIC.

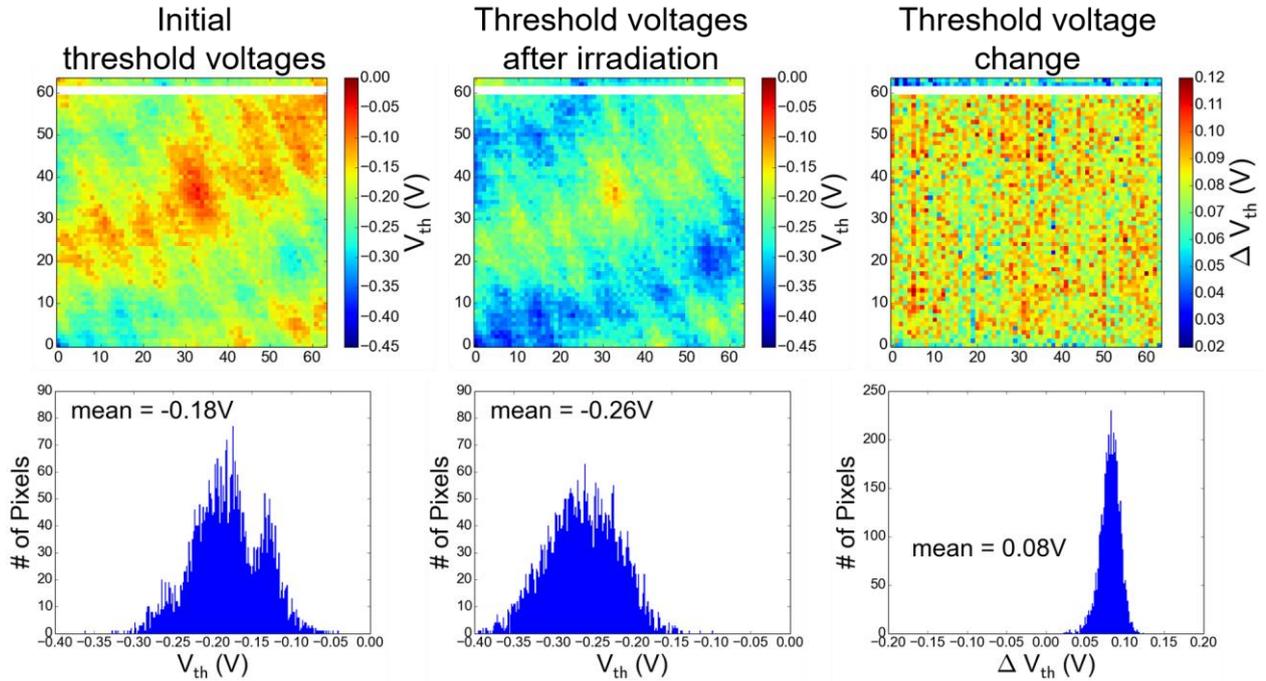

**Figure 6:** Threshold voltages of all 4096 DEPFETs measured before and after irradiation to a dose of 14 Gy

It is, however, important to factor in the case of inhomogeneous dose because a magnetic diverter for soft protons is planned for the WFI. Even if it is considered that a DEPFET is a back side illuminated sensor whose 450 µm thick bulk shields the susceptible front side from low energy charged particles (protons below 7 MeV cannot reach the front side and will not contribute to the TID), it is currently unclear whether the magnetic diverter could lead to an inhomogeneous dose distribution. With the results obtained in the TID test it is possible to estimate the consequences of an inhomogeneous dose: A threshold voltage shift of 100 mV together with the transconductance of the DEPFET ($g_m \approx 100$ µA/V) is giving rise to current changes of ~ 10 µA for pixels that receive the full dose of 14 Gy. As the initial current distribution of a large 512×512 pixel DEPFET has a width of ~60 µA and the input range of the I2V stage in the VERITAS is up to 120 µA [16] this is giving confidence that even in the case of very inhomogeneous irradiation the spread of the individual pixel currents will stay within acceptable limits.

A change in the readout noise of roughly 10% was observed. This result is a valuable input for further radiation tests, i.e. proton testing for the determination of leakage current increase due to displacement damage.

Even though the test was focused on the radiation response of the DEPFET sensor, the current versions of the ASICs have proven their radiation hardness up to the expected dose, further radiation studies (incl. SEE) and standalone tests are planned for the new versions currently under development.

## ACKNOWLEDGMENTS

Development and production of the DEPFET sensors for the Athena WFI are performed in a collaboration between MPE and the MPG Semiconductor Laboratory (HLL). We thank Olaf Hälker (MPE electronics), Rafael Strecker (MPE mechanics) and all other WFI team members for their contribution to the test.
The work was funded by the Max-Planck-Society and the German space agency DLR (FKZ: 50 QR 1901).